\documentclass{article}

\usepackage{amsmath,amscd}

\usepackage{epsfig}

\usepackage{amssymb}
\usepackage[usenames,dvipsnames]{color}

\usepackage{latexsym}

\usepackage{tikz}

\usepackage{hyperref}
 
\newtheorem{theorem}{Theorem}[section]
\newtheorem{prop}{Proposition}[section]
 \newtheorem{lemma}[theorem]{Lemma}

\newcommand{\mbf}{{\mathbb  F}}

\newcommand{\mpm}{{\mathcal P}M}

\newcommand{\oab}{\omega_{(A,B)}}

\newcommand{\ovA}{{\overline{A}}}

\newcommand{{\tlg}}{\tilde\gamma }

\newcommand{{\tlG}}{\tilde\Gamma }

\newcommand{\pap}{{\mathcal P}_{\ovA}P}

\newcommand{\mfc}{\mathbb {F}({\mathbf C})}

\newcommand{\mbc}{{\mathbf C}}

\newcommand{\Mor}{{\rm Mor}}

\newcommand{\End}{{\rm End}}

\usepackage{diagrams}

\begin{document}
 \title{A Morphism Double Category and Monoidal Structure}



\author{Saikat Chatterjee\\
School of Mathematics,\\
 Department of Mathematics,\\
 Harish Chandra Research Institute\\
Chhatnag Road,Jhusi,Allahabad-211 019,\\  
Uttar Pradesh, INDIA \\
email: saikatchatterjee@hri.res.in 
\and Amitabha Lahiri\\ S.~N.~Bose National Centre for Basic
Sciences \\ Block JD, Sector III, Salt Lake, Kolkata 700098 \\
  West Bengal, INDIA \\
 email: amitabha@bose.res.in
 \and
  Ambar N. Sengupta\\  Department of Mathematics\\
  Louisiana State University\\  Baton
Rouge, Louisiana 70803, USA\\
email:   ambarnsg@gmail.com 
 }

\maketitle

\begin{abstract}  A double category is constructed from a `fattened'
  version of a given category, motivated in part by a 
   context of parallel transport. We also study monoidal structures on the underlying
  category and on the fattened category.
  
   \end{abstract}



\section{Introduction and geometric background}\label{intro}

The interaction of point particles through a gauge field can be
encoded by means of Feynman diagrams, with nodes representing
particles and directed edges carrying an element of the gauge group
representing parallel transport along that edge. If the point
particles are replaced by extended one dimensional string-like
objects then the interaction between such objects can be encoded
through diagrams of the form
  \begin{equation}\label{D:xiyifidgm}\begin{diagram}
  & x_1 & \rTo^{f_1} & y_1 &\\
  &\dTo_{g_1} & \dImplies~h & \dTo_{g_2}&\\
  & x_2 &\rTo_{f_2} & y_2 &
  \end{diagram}\end{equation}
where the labels $f_i$ and $g_i$ describe  classical parallel
transport and $h$, which may take values in a different gauge
group, describes parallel transport over a space of paths.

 We will now give a rapid account of some of the geometric background.  We refer to our previous work   \cite{CLS2}
for further details. This material is not logically necessary for reading the rest of this paper, but is presented to indicate the context and motivation for some of the ideas of this paper.

Consider a principal $G$-bundle $\pi:P\to M$, where
$M$ is a smooth finite dimensional manifold and $G$ a Lie group,
and a connection $\ovA$ on this bundle. In the physical context,
$M$ may be spacetime, and $\ovA$ describes a gauge field. Now
consider the set $\mpm$ of piecewise smooth paths on $M$, equipped
with a suitable smooth structure.  Then the space $\pap$ of
$\ovA$-horizontal paths in $P$ forms a principal $G$-bundle  over
$\mpm$. We also use a second gauge group $H$ (that governs parallel transport over pathspace), which is a Lie group along with  a fixed   smooth homomorphism
$\tau:H\to G$ and a smooth map
$$G\times H\to
H:(g,h)\mapsto\alpha(g)h$$ such that each $\alpha(g)$ is an
automorphism of $H$, such that  
\begin{equation}\label{pfid}
\begin{split}
  \tau\bigl(\alpha(g)h\bigr) &=g\tau(h)g^{-1}\\
  \alpha\bigl(\tau(h)\bigr)h' &=hh'h^{-1}\end{split}
\end{equation}
 for all $g\in G$ and $h,h'\in H$. We denote the derivative  $\tau'(e)$ as $ \tau$, viewed as a map $LH\to LG$, and denote
  $\alpha'(e)$   by $\alpha$, to avoid notational complexity.  Given also a second connection form $A$ on $P$, and a smooth $\alpha$-equivariant vertical $LH$-valued $2$-form $B$ on $P$, it is possibly to construct a connection form $ \omega_{(A,B)}$ on the bundle $\pap$ \begin{equation}\label{def:omegaAB}
  \omega_{(A,B)}= {\rm ev}_1^*A +{\tau}(Z),
\end{equation}
where $Z$ is the $LH$-valued $1$-form on $\pap$ specified by
\begin{equation}\label{E:ZchenB}
Z=\int_0^1B,\end{equation}
which is a Chen integral. 

Consider a path of paths in $P$ specified through a smooth map
 $${\tilde\Gamma}:[0,1]^2\to
P:(t,s)\mapsto\tlG(t,s)=\tlG_s(t)=\tlG^t(s),$$ 
where each
$\tlG_s$ is $\ovA$-horizontal, and the path $s\mapsto\tlG(0,s)$
is $A$-horizontal. Let $\Gamma=\pi\circ\tlG$.   The {\em bi-holonomy} $g(t,s)\in G$   is specified as follows:
parallel translate $\tlG(0,0)$ along $\Gamma_0|[0,t]$ by $\ovA$,
then up the path $\Gamma^t|[0,s]$ by $A$, back along
$\Gamma_s$-reversed by $\ovA$ and then down $\Gamma^0|[0,s]$ by
$A$; then the resulting point is
\begin{equation}\label{defbihol}
\tlG(0,0)g(t,s).
\end{equation}

The following result is  proved in (\cite{CLS2}):

\begin{theorem}\label{T:gb} Suppose $${\tilde\Gamma}:[0,1]^2\to
  P:(t,s)\mapsto\tlG(t,s)=\tlG_s(t)=\tlG^t(s)$$ is smooth, with
  each $\tlG_s$ being $\ovA$-horizontal, and the path
  $s\mapsto\tlG(0,s)$ being $A$-horizontal.  Then the parallel
  translate of $\tlG_0$ by the connection $\oab$ along the path
  $[0,s]\to \mpm:u\mapsto\Gamma_u$, where $\Gamma=\pi\circ\tlG$,
  results in
\begin{equation}\label{E:tilgbihol}
  \tlG_s g(1,s)\tau\bigl(h_0(s)\bigr),
\end{equation} 
with $g(1,s)$ being the `bi-holonomy' specified as in
(\ref{defbihol}), and $s\mapsto h_0(s)\in H$ solving the
differential equation
\begin{equation}\label{difeqnhs}
\frac{dh_0(s)}{ds}h_0(s)^{-1}=-{
  \alpha}\bigl(g(1,s)^{-1}\bigr)\int_0^1B\bigl( 
\partial_t\tlG(t,s),\partial_s\tlG(t,s)\bigr)
\,dt 
\end{equation}
with initial condition $h_0(0)$ being the identity in $H$.
\end{theorem}

\begin{figure}[h!]
\centering
 \begin{tikzpicture}[xscale=.5,yscale=.5]
 
    \node [label= above: {$c_1$}] (c1) at
(0,3 ){};
 \node [label= below: {$c_2$}] (c2) at
(0,-2 ){};

\draw [fill] (-2,2) circle (.15cm);

 \node [label=above :{$E_{p_1}$}] (x1) at
(-2, 2) {};

\draw [fill] (2,2) circle (.15cm);
 \node [label=above :{$E_{q_1}$}] (y1) at
(2,2) {};

\draw [fill] (-2,-2) circle (.15cm);

 \node [label=below :{$E_{p_2}$}] (x2) at
(-2,-2) {};

\draw [fill] (2,2) circle (.15cm);

 \node [label=below :{$E_{q_2}$}] (y2) at
(2,-2) {};

\draw[->] (0,2.75)--(0.1,2.75);

\draw [thick] (-2,2) .. controls (-1,3) and (1,3) .. (2,2);

\draw [->] (-2,-2)--(0,-2)  ;

\draw (0,-2)--(2,-2);
\draw [->]  (-2,2)--(-2,0);
 
\draw  (-2,2)--(-2,-2);
\draw  (-2,-2)--(-2,0);

\draw   (2,-2)--(0,-2)  ;
 
\draw [->]  (2,2)--(2,0);
 \draw   (2,0)--(2,-2);
\draw  (2,2)--(2,-2);

      \end{tikzpicture}
      \caption{Paths and fibers}
      \label{F:pathfibers} 
      \end{figure}
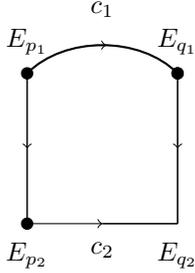

      Consider the category  $\mbc_0$ whose objects are fibers of a given vector bundle $E$ over $M$
      and whose arrows are piecewise smooth paths in $M$ (up to `backtrack equivalence'; for more on this notion see \cite{Le10}) along with parallel transport operators, by a connection $\ovA$, along such paths.  Note that all arrows are invertible.  In  Figure \ref{F:pathfibers}, $E_{p_1}$ is the vector space which is the fiber over the corresponding point  $p_1$. For the path $c_1$  there is a parallel transport operator $f_1:E_{p_1}\to E_{q_1}$. Next, if $c_2$ is a path from the base of the fiber $E_{p_2}$ to the base of $E_{q_2}$ then there is a corresponding parallel transport operator $f_2: E_{p_2}\to E_{q_2}$.

       A `higher' morphism $ c_1\to c_2$   is obtained from
      any suitably smooth path of paths, starting with the initial path $c_1$ and ending with $c_2$ (again backtracks need to be erased).  Using the connection $\ovA$   this produces parallel transport operators and paths $E_{p_1}\to E_{p_2}$ and $E_{q_1}\to E_{q_2}$. Moreover, another connection $A$ and  $2$-form $B$, along with a path of paths  leads to a linear map ${\rm Mor}_l(E_{p_1},E_{q_1})\to {\rm Mor}_l(E_{p_2},E_{q_2})$, where $\Mor_l(E,F)$ is  the vector space of all linear maps $E\to F$.  We  view this, in a `first approximation', as a morphism from the object ${\rm Mor}(E_{p_1},E_{q_1})$ to the object  $ {\rm Mor}(E_{p_2},E_{q_2})$ (say, mapping all paths from $p_1$ to $q_1$ to the path $c_2$).  In this paper we will not develop this framework in full detail (that would build on the theory from our earlier work    \cite{CLS2}) but focus on more algebraic aspects and other purely algebraic issues (such as monoidal structures).

Instead of vector bundles one could also work with the principal bundle  $P$ itself, taking as objects of a category $\mbc_0$
all the fibers of the bundle $P$, and as morphisms $f:P_{p}\to P_q$ the
$G$-equivariant bijections $P_p\to P_q$, where $P_p$ and $P_q$ are fibers
of $P$,  over points $p$ and $q$, and paths running from $p$ to $q$.
 
The interface between gauge theory and category theory, in various
forms and cases, has been studied in many works, for instance
\cite{Baez, BS, BS2, CLS2, GP, Lahiri}. In the present paper, we extract 
the abstract essence of some of
these structures in a category theory setting, leaving the
differential geometry behind as the concrete context.  We abstract
the process of passing from the point-particle picture to a
string-like picture to a functor which generates a category $\mfc$
from a category $\mbc$.  Our  Proposition
\ref{P:fatmon}, describes properties of a natural product operation
on the objects of $\mfc$ when $\mbc$ is a monoidal category.    An excellent
review of monoidal categories in relation to topological
quantum field theory can be found in \cite{Baez2}. Symmetric 
monoidal bicategories are discussed in \cite{shul} in a context 
different from ours.

\section{The Fat Category}\label{S:fc}

Let $\mathbf C$ be a category. We define a new category $\mathbb
{F}({\mathbf C})$ as follows. The objects of $\mfc$ are the
morphisms of $\mbc$.  A morphism in $\mfc$ from the object
$x_1\stackrel{f _1}{ \rightarrow} y_1$ to the object
$x_2\stackrel{f _2}{ \rightarrow} y_2$ consists of morphisms
$x_1\stackrel{g _1}{ \rightarrow} x_2$ and $y_1\stackrel{g_2}{
  \rightarrow} y_2$ in $\mathbf C$, along with a set-mapping
\begin{equation}\label{E:hdef}
h: {\rm Mor}(x_1,y_1)\to {\rm Mor}(x_2,y_2)
\end{equation}
which maps $f_1$ to $f_2$:
$$h(f_1)=f_2.$$ 
(In  a later section we require that the hom-sets ${\rm Mor}(x,y)$ themselves also have algebraic structure that should be preserved by such $h$.) 
Here is a diagram displaying a morphism $u$ of $\mfc$:
\begin{equation}  \label{E:f1g1h}
u\quad =\quad\begin{diagram} 
x_1 &\rTo^{f_1} & y_1 \\ 
\dTo^{g_1} & \dImplies~h & \dTo_{g_2}\\
x_2 & \rTo_{f_2} & y_2
 \end{diagram}
 \end{equation} 
It is clear that this  does specify  a category, which we call the
{\em fat category} for $\mbc$ (composition is `vertical', with successive $h$'s composed).  Sometimes it will be easier on the
eye to write 
$$(x,y,f)$$
for $x\stackrel{f}{\rightarrow} y$. Thus, the diagram
(\ref{E:f1g1h}) can also be displayed as 
\begin{equation}  \label{E:f1g1hsh}
\begin{diagram} 
(x_1, y_1,f_1) \\ 
  \dTo_u  \\
(x_2, y_2,f_2)
 \end{diagram}
 \end{equation}
The composition $v\circ_V u$ of morphisms in $\mfc$ is defined
`vertically' by drawing the diagram of $v$ below that of $u$ and
composing vertically downward.   

Commutative diagrams in ${\mathbf C}$ lead to morphisms of $\mfc$ in a natural way, and yield a subcategory of $\mfc$ that is recognizable as the `category of arrows' \cite[\S I.4]{MacLane} sometimes denoted ${\rm Arr}({\mathbf C})$. 

\begin{lemma}\label{L:induce}
Any commutative diagram 
\begin{displaymath}  
\begin{diagram} 
x_1 &\rTo^{f_1} & y_1 \\ 
\dTo^{g_1} &  & \dTo_{g_2}\\
x_2 & \rTo_{f_2} & y_2
 \end{diagram}
\end{displaymath}
in $\mbc$, in which $g_1$ is an isomorphism,  generates a morphism 
$$(x_1,y_1,f_1)\stackrel{u}{\rightarrow} (x_2,y_2,f_2)$$
in $\mfc$,
\begin{displaymath}  
u\qquad=\qquad \begin{diagram} 
x_1 &\rTo^{f_1} & y_1 \\ 
\dTo^{g_1} & \dImplies~{h_u} & \dTo_{g_2}\\
x_2 & \rTo_{f_2} & y_2
 \end{diagram}
\end{displaymath}
where 
\begin{equation}\label{E:hph}
h_u: {\rm Mor}(x_1,y_1)\to {\rm Mor}(x_2,y_2): \phi\mapsto
g_2{\phi}g_1^{-1}.\end{equation} 
Moreover, if 
\begin{displaymath}  
 \begin{diagram} 
x_1 &  \rTo^{f_1} & y_1   \\ 
\dTo^{g_1 }  &  & \dTo_{g_2} \\
x_2 & \rTo_{f_2} & y_2 \\
\dTo^{g'_1 }  &  & \dTo_{g'_2} \\
x_3 & \rTo_{f_3} & y_3 \\
 \end{diagram}
\end{displaymath}
is a commutative diagram in $\mbc$, where $g_1$ and $g'_1$ are
isomorphisms, then the  composite of the induced morphisms 
$$u:(x_1,y_1,f_1)\to (x_2,y_2,f_2)\quad\hbox{and}\quad
v:(x_2,y_2,f_2)\to (x_3,y_3,f_3)$$ 
is the morphism in $\mfc$  induced by the commutative diagram
\begin{displaymath}  
\begin{diagram}
x_1 &\rTo^{f_1}   & y_1\\ 
\dTo^{g _2g_1} &  &\dTo_{g'_2g'_1}\\
x_3 & \rTo_{f_3}  &z_3
 \end{diagram}
\end{displaymath}
\end{lemma}

  \section{A Double Category of Isomorphisms}\label{S:dc}
  
Let $\mfc_0$ be the category whose objects are the invertible arrows of
$\mbc$    and whose arrows are the arrows   
 \begin{equation}\label{D:x1g1f1h}\begin{diagram}
  & x_1 & \rTo^{f_1} & y_1 &\\
  &\dTo^{g_1} & \dImplies~h & \dTo_{g_2}&\\
  & x_2 &\rTo^{f_2} & y_2 &
  \end{diagram}\end{equation}
  in $\mfc$ in which the verticals $g_1$ and $g_2$ are also
  isomorphisms in $\mbc$.  This is, for all purposes here, as good as assuming
  that all arrows of $\mbc$
  are invertible, since henceforth we will only work with such arrows. In the geometric context, the arrows represent parallel transports and so the invertibility assumption
  is natural. The mapping $h$ is motivated by the `surface' parallel-transport mentioned briefly in (\ref{difeqnhs}).
  
Let us define 
{\em horizontal composition} of morphisms in
 $\mfc_0$ as follows:
  \begin{equation}  \label{E:horcomp}
 \begin{diagram}  
x_1 &\rTo^{f_1} & y_1 \\ 
\dTo^{g_1} & \dImplies~h & \dTo_{g_2}\\
x_2 & \rTo_{f_2} & y_2
 \end{diagram}\quad\circ_{H} \quad \begin{diagram}  
y_1 &\rTo^{f'_1} & z_1 \\ 
\dTo^{g'_1} & \dImplies~h' & \dTo_{g'_2}\\
y_2 & \rTo_{f'_2} & z_2
 \end{diagram}\quad=\quad \begin{diagram}  
x_1 &\rTo^{f'_1f_1} & z_1 \\ 
\dTo^{g_1} & \dImplies~{h''} & \dTo_{g'_2}\\
x_2 & \rTo_{f'_2f_2} & z_2
 \end{diagram}
 \end{equation}
 where the composition is defined only when $g'_1=g_2$,  and $h''$
 is given by 
\begin{equation}\label{E:h'h}
h'':{\rm Mor}(x_1,z_1)\to {\rm Mor}(x_2,z_2): f \mapsto
h'(ff_1^{-1})  \underbrace{h(f_1) }_{f_2}\end{equation} 
Note that $h''$ is  satisfies
\begin{equation}\label{E:hdp}
h''(f'_1f_1)=
h'(f'_1)h(f_1)=f_2'f_2.\end{equation}
Consider now the following diagram:
\begin{equation}  \label{E:window1}
 \begin{diagram}  
x_1 &\rTo^{f_1} & y_1 & \rTo^{f'_1} & z_1\\
\dTo^{g_1} &\dImplies~h  &\dTo_{g_2} & \dImplies~{h'} & \dTo_{g_3}\\
x_2 & \rTo_{f_2} & y_2 & \rTo_{f'_2} & z_2\\
\dTo^{g'_1} &\dImplies~j  &\dTo_{g'_2}  & \dImplies~{j'} & \dTo_{g'_3}\\
x_3 & \rTo_{f_3} & y_3 & \rTo_{f'_3} & z_3\\
 \end{diagram}
 \end{equation}

The morphisms of  $\mfc_0$ thus have two   laws of
composition $\circ_V$ and $\circ_H$. As we see below, these compositions obey a consistency condition  (\ref{E:dclaw}), which thereby specifies a {\em double
category} \cite{KS} and \cite[\S I.5]{MacLane}.

\begin{prop}\label{P:dc} The morphisms of $\mfc_0$ form a double category under the laws of composition $\circ_V$ and $\circ_H$ in the sense that for  the diagram (\ref{E:window1}),   with
  notation as   explained above,  
 \begin{equation}\label{E:dclaw}
(u_{j'}\circ_H u_j) \circ_V (u_{h'}\circ_H u_h)=   (u_{j'}\circ _Vu_{h'})\circ_H(u_j\circ_V
 u_h), 
 \end{equation}
  for all morphisms $u_{j'}$, $u_j$, $u_h$, $u_{h'}$ in ${\rm Mor}(\mfc_0)$ for which the compositions  on both sides of   (\ref{E:dclaw}) are meaningful.  \end{prop}
 \noindent\underline{Proof}.  Denote by $u_{h}$ the morphism of $\mfc_0$ specified by the upper left square in (\ref{E:window1}), by $u_{h'}$ the morphism specified by the upper right square, by $u_j$ the morphism specified by the lower left square and, lastly, by $u_{j'}$ the morphism specified by the lower right square.
 
 Let $f\in {\rm Mor}(x_1, z_1)$. Then
 \begin{equation}\label{E:hv1}
 \begin{split}
\left((u_{j'}\circ_H u_j) \circ_V (u_{h'}\circ_H u_h)\right)(f) &=((u_{j'}\circ_H u_j) (h'(ff_1^{-1})f_2)\\
&= {j'}\bigl(h'(ff_1^{-1})\bigr)f_3, \end{split}\end{equation}
and 
\begin{equation}\label{E:hv2}\begin{split}
\left( (u_{j'}\circ _Vu_{h'})\circ_H(u_j\circ_V
 u_h)\right)(f) &=\bigl((u_{j'}\circ_V u_{h'}) (ff_1^{-1})\bigr) f_3 \\
&= {j'}\bigl(h'(ff_1^{-1})\bigr)f_3.\end{split}\end{equation}
Comparing  (\ref{E:hv1}) and  (\ref{E:hv2}), we have the claimed equality (\ref{E:dclaw}).
\fbox{QED}
 
Then $\mfc_0$ equipped with both laws of
composition $\circ_V$ and $\circ_H$ is a {\em double
category} \cite{KS}. In the geometric context this is expressed
as a {\em flatness} condition for the connection
$\omega_{\ovA,A,B}$ described in the Introduction; for more, see,
for instance, \cite{Baez, CLS2}. 

\section{Enrichment for Morphisms}\label{S:enrich}

In the geometric context  the objects of $\mbc$ are real vector spaces, fibers of some vector bundle. A morphism $x\to y$ is specified by a path, from the base of $x$ to the base of $y$, along with a linear map $x\to y$, representing parallel transport along the path. Extracting the abstract essence of this, let  $\mfc_1$  be a category whose objects are the hom-sets of $\mbc$.  The idea is that the objects in $\mfc_1$ could have additional structure; for example, if $\mbc$ has only one object $E_p$, a fiber of a vector bundle, then $\Mor(E_p,E_p)$ is a group under composition. The  category  $\mfc_1$ is a subcategory of $\mfc_0$, having the same objects but possibly fewer morphisms; in the example, the morphisms of $\mfc_1$ could be required to be group automorphisms. We require that for any objects $x,y,z$ of $\mbc$  and isomorphism $g:y\to x $,  the map
\begin{equation}\label{E:mrxyzfmor}
r_g: \Mor(x,z)\to\Mor(y,z): f\mapsto f g\end{equation}
be a morphism of $\mfc_1$.

  \begin{prop}\label{P:fc0cat}    Let $\mfc_1$ be any subcategory of $\mfc_0$ having the same objects as $\mfc_0$, and satisfying the condition (\ref{E:mrxyzfmor})  as explained above. Both horizontal and vertical composites of morphisms in $\mfc_1$ are in
    $\mfc_1$.  Thus,
    $\mfc_1$ is a double category.  \end{prop} 
    \noindent\underline{Proof}.  The consistency condition between horizontal and vertical compositions has already been checked in Proposition \ref{P:dc}. Thus we need only check that horizontal composition, specified in (\ref{E:h'h}) as 
    \begin{equation}\label{E:h'h2}
h'':{\rm Mor}(x_1,z_1)\to {\rm Mor}(x_2,z_2): f \mapsto
h'(ff_1^{-1})   h(f_1),\end{equation} 
is a morphism of  $\mfc_1$,  for all  invertible $f_1\in \Mor(x_1, y_1)$, and  all $h:\Mor(x_1,y_1)\to\Mor(x'_1,y'_1)$, $h'\in\Mor(y_1,z_1)\to\Mor(y'_1,z'_1)$   morphisms in $\mfc_1$. Observe that
\begin{equation}
h''(f)=h'(ff_1^{-1})   h(f_1)=  r_{h(f_1)} \circ h'\circ r_{f_1^{-1}} (f)\end{equation}
where the notation $r_g$ is as in (\ref{E:mrxyzfmor}). Thus $h''$ is a composite of morphisms in  $\mfc_1$.   \fbox{QED}
 
\section{Monoidal structures}\label{S:monoidal}

In this section we will explore some algebraic structural enhancements of the fattened category $\mfc_0$.  The discussion is motivated by intrinsic algebraic considerations, but we discuss briefly now the relationship with the geometric context.

Consider the very special case where $\mbc$ is the category with only one object $E_o$, the fiber over a fixed point $o$ in a vector bundle, and a morphism $f:E_o\to E_o$ is a an ordered pair:
$$f=(c, T),$$
  consisting of a piecewise smooth loop  $c$ based at $o$ (with backtracks erased) along with a linear map $T:E_o\to E_o$ representing parallel transport around the loop.  For $\mfc_0$ in this special case, a morphism $h:\Mor(E_o,E_o)\to \Mor(E_o,E_o)$ is a {\em path of paths}  along with a linear map $\End(E_o)\to\End(E_o)$, were $\End(E_o)$ is the vector space of all linear maps $E_o\to E_o$. Each hom-set $\Mor(E_o,E_o)$ is a monoid: composition 
$$\Mor(E_o,E_o)\times\Mor(E_o,E_o)\to\Mor(E_o,E_o):(f,f')\mapsto f\otimes f'$$
is given simply by concatenation  of loops along with ordinary composition of linear maps in $\End(E_o)$:
$$(c,T)\otimes (c',T')=(c\ast c', T\circ T'),$$
where $c\ast c'$ is the loop $c'$ followed by the loop $c$.
 (Since this discussion is primarily for motivation, we leave out  techincal details of   `backtrack erasure'.)

Turning to the abstract setting, we assume henceforth that    $\mbc$ is a monoidal category. This means that
there is a bifunctor 
$$\otimes:\mbc\times\mbc\to\mbc$$
and there is an {\em identity object} 
$1$ in $\mbc$ for which certain natural coherence 
conditions hold as we now describe. In addition, there exists a natural isomorphism
$\alpha$, the {\em associator}, which associates to any objects $A$,
$B$, $C$ of $\mbc$ an isomorphism 
$$\alpha_{A,B,C}:(A\otimes B)\otimes C\to A\otimes (B\otimes C)$$
such that the following diagram commutes:
{\scriptsize
\begin{displaymath}  
\begin{diagram} 
((A\otimes B)\otimes C)\otimes D &\rTo^{\alpha_{A,B,C}\otimes i_D}
&(A\otimes (B\otimes C))\otimes D &\rTo^{\alpha_{A,B\otimes C,D} }
& A\otimes ((B\otimes C)\otimes D)\\  
\dTo^{\alpha_{A\otimes B,C,D}} &&&& \dTo_{i_A\otimes
  \alpha_{B,C,D}}  \\ 
(A\otimes B)\otimes (C\otimes D)  & &\rTo_{\alpha_{A,B,C\otimes D}}
&&A\otimes (B\otimes (C\otimes D)) 
\end{diagram}
\end{displaymath}
}
There are also natural isomorphisms $l$ and $r$, the left and right
{\em unitors}, associating to each object $A$ in $\mbc$ morphisms 
$$l_A:1\otimes A\to A\qquad\hbox{and}\qquad r_A:A\otimes 1\to A$$
such that
\begin{displaymath}  
\begin{diagram} 
(A\otimes 1)\otimes B  &&\rTo^{\alpha_{A,1,B}} & &A\otimes (1\otimes B)\\
& \rdTo_{r_A\otimes i_B} &&  \ldTo_{i_A\otimes l_B} &\\
&& A\otimes B &&
\end{diagram}
\end{displaymath}
commutes for all objects $A$ and $B$ in $\mbc$. 

Note that {\em naturality} means there are certain other conditions
as well. For example, that the left unitor is a natural
transformation means that for any  morphism
$x\stackrel{f}{\rightarrow} y$ in $\mbc$ the diagram 
\begin{equation}  \label{E:lunat}
\begin{diagram} 
 1\otimes x & \rTo^{1\otimes f} & 1\otimes y\\
 \dTo_{l_x} && \dTo^{l_y}\\
 x&\rTo^f&y
\end{diagram}
\end{equation}
commutes; here, in the upper horizontal arrow, $1$ is the unique
morphism $i_1:1\to 1$ in $\mbc$.

We now define a product on the objects of $\mfc$:
$${\rm Obj}(\mfc)\times {\rm Obj}(\mfc)\to  {\rm Obj}(\mfc):(u,v)\mapsto u\otimes v$$
as follows
\begin{equation}\label{E:tensfc}
(x_1\stackrel{f_1}{\rightarrow} y_1)\otimes
(x_2\stackrel{f_2}{\rightarrow} y_2)\, 
\stackrel{\rm def}{= }\,x_1\otimes x_2\stackrel{f_1\otimes
  f_2}{\rightarrow} y_1\otimes y_2 
\end{equation}

In the fat category $\mfc$ we then have associators and unitors as
follows.  First, the unit is 
$$1_{\mbf}=1\stackrel{i_1}{\rightarrow} 1$$
where $1$ denotes   the identity object in $\mbc$ and $i_1$ the
identity map on $1$. We will often denote   $i_1$   also simply as
$1$, the meaning being clear from context. 
For any object $x\stackrel{f}{\rightarrow} y$  there is the left unitor
\begin{equation}\label{D:leftunit}
\begin{diagram}
  1_{\mbf}\otimes (x,y,f)&& 1\otimes x & \rTo^{1\otimes f} &
  1\otimes y\\ 
\dTo^{l^\mbf_{(x,y,f)}} & \qquad = \qquad &\dTo^{l_x} &
\dImplies~{l_{(x,y,f)}} & \dTo_{l_y}\\ 
(x,y,f) && x &\rTo_{f} & y
\end{diagram}
\end{equation}
 where the mapping
$$l_{(x,y,f)}:{\rm Mor}(1\otimes x, 1\otimes y)\to {\rm Mor}(x,y):
\phi\mapsto l_y{\phi}l_x^{-1}$$
  takes $1\otimes f$ to $f$, as follows from the remarks made above
  for (\ref{E:lunat}). 
The  right unitor   is

\begin{equation}\label{D:rightunit}
\begin{diagram}
  x\otimes 1 & \rTo^{f\otimes 1} & y\otimes 1\\
\dTo^{r_x} & \dImplies~{r_{  (x,y,f) } }& \dTo_{r_y}\\
x &\rTo_{f} & y
\end{diagram}
\end{equation}
where
$$r_{  (x,y,f)}:{\rm Mor}(x\otimes 1, y\otimes 1)\to {\rm Mor}(x,y):
\phi\mapsto r_y{\phi}r_x^{-1}$$
Again this is indeed a morphism in $\mfc$ by essentially the same
argument as was used above in (\ref{D:leftunit}) for the left unitor.

The associator in $\mfc$ is given as follows.  Consider objects
$x_i\stackrel{f_i}{\rightarrow}y_i$ in $\mfc$, for
$i\in\{1,2,3\}$. The fact that $\alpha$ is a {\em natural}
transformation means that the diagram 
\begin{equation}\label{E:prodassoc}
\begin{diagram} 
(x_1\otimes x_2)\otimes x_3 & \rTo^{(f_1\otimes f_2)\otimes f_3} & (y_1\otimes y_2)\otimes y_3\\
\dTo^{\alpha_{x_1,x_2,x_3}} &&  \dTo_{\alpha_{y_1,y_2,y_3}}\\
x_1\otimes (x_2\otimes x_3)  & \rTo_{f_1\otimes (f_2\otimes f_3)} & y_1\otimes (y_2\otimes y_3)
\end{diagram}
\end{equation}
is commutative. Hence, by the first half of Lemma \ref{L:induce},
this induces a morphism 
{\scriptsize 
\begin{equation}\label{E:prodassoch}
\begin{diagram} 
 \bigl( (x_1,y_1,f_1)\otimes (x_2,y_2,f_2)\bigr)\otimes (x_3,y_3,f_3) \\
\dTo~h\\ 
 (x_1,y_1,f_1)\otimes   \bigl( (x_2,y_2,f_2) \otimes (x_3,y_3,f_3)\bigr)
\end{diagram}
\begin{diagram}\qquad =\qquad \end{diagram}
\begin{diagram} 
(x_1\otimes x_2)\otimes x_3 & \rTo^{(f_1\otimes f_2)\otimes f_3} &
(y_1\otimes y_2)\otimes y_3\\ 
  &\dTo~h&   \\
x_1\otimes (x_2\otimes x_3)  & \rTo_{f_1\otimes (f_2\otimes f_3)} &
y_1\otimes (y_2\otimes y_3) 
\end{diagram}
\end{equation}
}
in $\mfc$. In fact, $h$ is an isomorphism since the vertical arrows
in (\ref{E:prodassoc}) are isomorphisms. 

We prove the coherence condition for unitors. For this we have the diagram
 
\begin{equation*}\begin{diagram}
(x_1\otimes 1)\otimes x_2 & \rTo^{\alpha_{x_1,1,x_2}} & x_1\otimes
(1\otimes x_2) && &\\ 
& \rdTo(3,2) \rdTo(1,2) _{l_{x_1}\otimes 1_{x_2}} \ldTo(1,2)&    &
\rdTo(3,2)^{f_1\otimes (1\otimes f_2)}&\\ 
& x_1\otimes x_2 && (y_1\otimes 1)\otimes y_2
&\rTo^{\alpha_{y_1,1,y_2}} & y_1\otimes (1\otimes y_2)\\ 
 & & \rdTo(3,2)_{f_1\otimes f_2} & & \rdTo(1,2)\ldTo(1,2)& \\
& & & & y_1\otimes y_2 &
\end{diagram}
\end{equation*}
The two triangles at the two ends of this `trough' commute because
of coherence in $\mbc$,  the top rectangle also commutes because of
naturality of $\alpha$. Then it is entertaining to check that the
two rectangular `slanted sides' are also commutative.  In fact, the
slant side on the left is  
$$r^{\mbf}_{(x_1,y_1,f_1)}\otimes 1_{(x_2,y_2,f_2)}:\bigl(
(x_1,y_1,f_1)\otimes 1^{\mbf}\bigr)\otimes (x_2,y_2,f_2)\to
(x_1,y_1,f_1)\otimes (x_2,y_2,f_2)$$ 
as a morphism in $\mfc$, 
and the slant side on the right is 
$$1_{(x_1,y_1,f_1)}\otimes l^{\mbf}_{(x_2,y_2,f_2)}$$
 Thus, viewed as a diagram in $\mfc$, the `trough' looks like

\begin{equation}\label{E:troughav}
\begin{diagram} 
\bigl((x_1,y_1,f_1)\otimes 1_{\mbf}\bigr)\otimes (x_2,y_2,f_2) &
\rTo^{\alpha^\mbf} & (x_1,y_1,f_1)\otimes \bigl(1_{\mbf}\otimes
(x_2,y_2,f_2) \bigr)\\ 
&\rdTo(1,3)_{r^{\mbf}_{(x_1,y_1,f_1)}\otimes
  1_{(x_2,y_2,f_2)}}\ldTo(1,3)_{1_{(x_1,y_1,f_1)}\otimes
  l^{\mbf}_{(x_2,y_2,f_2)}} &\\ 
&&\\
& (x_1,y_1,f_1)\otimes (x_2,y_2,f_2)
\end{diagram}
\end{equation}
Since the trough commutes in $\mbc$ so does its avatar
(\ref{E:troughav}) in $\mfc$, 
thanks to  the second half of Lemma \ref{L:induce}.
This verifies the coherence property in $\mfc$ involving the unitors.

Now we turn to coherence for the associators. In the following 
diagram, where we leave out the $\otimes$ products for ease of 
viewing, the slant arrows are all tensor products of the $f_i$  
and the horizontal and vertical arrows are various associators: 
{\scriptsize
\begin{equation}\label{E:coherassoc}
\hskip -.2in
\begin{diagram} 
& \bigl((x_1  x_2) x_3\bigr) x_4&& \rTo& \bigl((x_1(x_2x_3)x_4\bigr) 
&\rTo & x_1\bigl((x_2x_3)\bigr)x_4 & & & &\\
& \dTo_{\alpha_{x_1x_2,x_3,x_4}} &   \rdTo(3,4) &  & & \rdTo(3,4) & \dTo & \rdTo(3,4)^{f_1((f_2f_3)f_4)}& &    &\\
& (x_1x_2)(x_3x_4) && \rTo & & & x_1\bigl(x_2(x_3x_4)\bigr) & & & & & &\\
& & \rdTo(3,4)_{(f_1f_2)(f_3 f_4)} & & & &&\rdTo(3,4) & & &   &&
\\
& & & &    \bigl((y_1  y_2) y_3\bigr)y_4 &&\rTo &\bigl((y_1(y_2y_3)\bigr)y_4&  \rTo & y_1\bigl((y_2y_3)y_4\bigr)&& \\
& & & &  \dTo & & & &&  \dTo  \\
& & & &  (y_1y_2)(y_3y_4) &  &\rTo^{\alpha_{y_1,y_2,y_3y_4}}   &   & &y_1\bigl(y_2(y_3y_4)\bigr) 
\end{diagram}
\end{equation}
}

Coherence in the monoidal category $\mbc$ implies that the two
rectangles at the end of this box are commutative, as mentioned 
earlier. Naturality of the associator implies that the top, bottom 
and sides are also commutative. Thus the entire diagram is commutative.  
If we abbreviate the objects in $\mfc$ as 
$$X_i=(x_i,y_i,f_i),$$ for $i\in\{1,2,3,4\}$, we can read the full 
diagram as a  diagram in the category $\mfc$: 
\begin{equation}\label{E:coherassocmfc}
\begin{diagram} 
& \bigl((X_1X_2)X_3\bigr)X_4& \rTo & \bigl((X_1(X_2X_3)\bigr)X_4
&\rTo & X_1\bigl((X_2X_3)\bigr)X_4 &\\ 
&\dTo & & && \dTo\\
& (X_1X_2)(X_3X_4)& \rTo &   &  &  X_1\bigl(X_2(X_3X_4)\bigr) &\\
\end{diagram}
\end{equation}
As a diagram in $\mfc$ this is commutative, by Lemma
\ref{L:induce}. This establishes coherence of the associator in
$\mfc$.  

We have completed the proof of

\begin{prop}\label{P:fatmon} Suppose $\mbc$ is a monoidal
  category and  let $\mfc$ be the category specified above in the
  context of (\ref{E:hdef}). Then, with tensor product as defined
  in (\ref{E:tensfc}), $\mfc$ satisfies all conditions of a
  monoidal category at the level of objects. 
\end{prop}

{\bf Acknowledgments.}  AL acknowledges research support from
Department of Science and Technology, India under Project
No.~SR/S2/HEP-0006/2008. ANS acknowledges research supported from
US NSF grant DMS-0601141.  A highly effective set of macros
developed by Paul Taylor has been used to draw all diagrams.

\end{document}